\documentclass[cits]{PoS}

\usepackage{amsmath,amssymb,amsfonts}

\providecommand{\nt}{\notag}
\providecommand{\ec}{\;,}
\providecommand{\ep}{\;.}
\providecommand{\tn}[1]{\textnormal{#1}}
\providecommand{\pipi}{$\pi\pi$~}
\providecommand{\piN}{$\pi N$~}
\providecommand{\sch}{$\pi N\to\pi N$}
\providecommand{\tch}{$\pi\pi\to\bar{N}N$}
\providecommand{\diff}{\!\!\tn{d}}
\providecommand{\mN}{m}
\providecommand{\mNs}{m^2}
\providecommand{\mpi}{M_\pi}
\providecommand{\mpis}{M_\pi^2}
\providecommand{\tm}{t_\tn{m}}
\providecommand{\tpi}{t_\pi}
\providecommand{\tN}{t_N}

\providecommand{\GeV}{\,\tn{GeV}}
\providecommand{\Ord}{\mathcal{O}}

\renewcommand{\Im}{\text{Im}\,}

\title{Roy--Steiner equations for $\boldsymbol{\pi N}$ scattering}
\ShortTitle{Roy--Steiner equations for \piN scattering}

\author{\speaker{Christoph Ditsche},$^a$ Martin Hoferichter,$^{ab}$ Bastian Kubis,$^a$ and Ulf-G.~Mei\ss ner $^{ac}$\\
        \llap{$^{a}$}Helmholtz-Institut f\"ur Strahlen- und Kernphysik (Theorie) and Bethe Center for Theoretical Physics, Universit\"at Bonn, D-53115 Bonn, Germany\\
        \llap{$^{b}$}Albert Einstein Center for Fundamental Physics, Institute for Theoretical Physics, Universit\"at Bern, CH-3012 Bern, Switzerland\\
        \llap{$^{c}$}Institut f\"ur Kernphysik, Institute for Advanced Simulation, and J\"ulich Center for Hadron Physics, Forschungszentrum J\"ulich, D-52425 J\"ulich Germany\\
        E-mail: \email{ditsche@hiskp.uni-bonn.de}}

\abstract{Starting from hyperbolic dispersion relations for the invariant amplitudes of pion--nucleon scattering together with crossing symmetry and unitarity, one can derive a closed system of integral equations for the partial waves of both the $s$-channel (\sch) and the $t$-channel (\tch) reaction, called Roy--Steiner equations.
After giving a brief overview of the Roy--Steiner system for \piN scattering, we demonstrate that the solution of the $t$-channel subsystem, which represents the first step in solving the full system, can be achieved by means of Muskhelishvili--Omn\`es techniques.
In particular, we present results for the $P$-waves featuring in the dispersive analysis of the electromagnetic form factors of the nucleon.}

\FullConference{The 7th International Workshop on Chiral Dynamics,\\
		August 6 -10, 2012\\
		Jefferson Lab, Newport News, Virginia, USA}

\begin{document}

\section{Introducing Roy--Steiner equations for \piN scattering}

Partial-wave dispersion relations (PWDRs) together with unitarity and crossing symmetry as well as isospin and chiral symmetry (i.e.\ all available symmetry constraints) have repeatedly proven to be a powerful tool for studying processes at low energies with high precision~\cite{ACGL,GKPY,BDM,HPS}.
For \piN scattering the (unsubtracted) hyperbolic dispersion relations (HDRs) for the usual Lorentz-invariant amplitudes read~\cite{HiteSteiner} (using the notation of~\cite{Hoehler}, see~\cite{DHKM} for more details)
\begin{align}
\label{eqn:HDRs}
A^+(s,t) &= \frac{1}{\pi}\int\limits_{s_+}^\infty\diff s'\left[\frac{1}{s'-s}+\frac{1}{s'-u}-\frac{1}{s'-a}\right]\Im A^+(s',t')
+\frac{1}{\pi}\int\limits_{\tpi}^\infty\diff t'\;\frac{\Im A^+(s',t')}{t'-t}\ec\nt\\
B^+(s,t) &= N^+(s,t)
+\frac{1}{\pi}\int\limits_{s_+}^\infty\diff s'\left[\frac{1}{s'-s}-\frac{1}{s'-u}\right]\Im B^+(s',t')
+\frac{1}{\pi}\int\limits_{\tpi}^\infty\diff t'\;\frac{s-u}{s'-u'}\frac{\Im B^+(s',t')}{t'-t}\ec\nt\\
N^+(s,t) &= g^2\left[\frac{1}{\mNs-s}-\frac{1}{\mNs-u}\right]\ec \qquad 
(s-a)(u-a)=b=(s'-a)(u'-a)\ec
\end{align}
and similarly for $A^-$, $B^-$, and $N^-$, where $N^\pm$ are the nucleon pole terms and the ``external'' (unprimed) and ``internal'' (primed) kinematics are related by real hyperbola parameters $a$ and $b$ (as well as via $s+t+u=2(\mNs+\mpis)=s'+t'+u'$), so that HDRs allow for the combination of all physical regions, which is known to be crucial for a reliable continuation into the subthreshold region and hence for an accurate determination of the \piN $\sigma$-term.
Furthermore, the imaginary parts are only needed in regions where the corresponding partial-wave decompositions converge and the range of convergence can be maximized by tuning the free hyperbola parameter $a$.
While the $s$-channel integrals start at the threshold $s_+=W_+^2=(\mN+\mpi)^2$, the $t$-channel contributes already above the pseudothreshold $\tpi=4\mpis$ far below the threshold $\tN=4\mNs$.
Depending on the asymptotic behavior of the imaginary parts, in principle it could be necessary to subtract the HDRs to ensure the convergence of the integrals, thereby parameterizing high-energy information with polynomials containing a priori unknown subtraction constants.
However, (additional) subtractions may also be introduced to lessen the dependence of the low-energy solution on high-energy input; the corresponding subtraction parameters then obey respective sum rules.
For \piN scattering it proves particularly useful to subtract at the subthreshold point $(s=u,t=0)$, as this preserves the $s\leftrightarrow u$ crossing symmetry (which can be made explicit in terms of the crossing variable $\nu=(s-u)/(4\mN)$ via $D^\pm(\nu,t)=A^\pm+\nu B^\pm=\pm D^\pm(-\nu,t)$).
This is especially favorable for the $t$-channel subproblem and facilitates matching to chiral perturbation theory~\cite{BuettikerMeissner,BecherLeutwyler} to determine the subtraction constants, which thus can be identified with the subthreshold expansion parameters.\footnote{For the PWDRs of \pipi scattering, called Roy equations~\cite{Roy}, an analogous matching procedure for the \pipi scattering lengths as pertinent subtraction parameters has been conducted in~\cite{CGL}. In contrast to $\pi\pi$ scattering, the $\pi N$ scattering lengths can be extracted with high accuracy from hadronic-atom data~\cite{Gotta+Strauch,piNcoupling} and may thus serve as additional constraints on the subtraction constants in the Roy--Steiner system.}
In addition to the presentation in~\cite{DHKM}, we also introduce a (partial) third subtraction, which is related to the parameters $a_{10}^+$ and $a_{01}^-$ of the subthreshold expansions (with $d_{0n}^+=a_{0n}^+$ for all $n\geq0$)
\begin{equation}
\label{eqn:p3sub}
A^+(\nu,t) = \frac{g^2}{\mN} +d_{00}^+ +d_{01}^+t +a_{10}^+\nu^2 +\Ord\big(\nu^4,\nu^2t,t^2\big)\ec \qquad 
A^-(\nu,t) = a_{00}^-\nu +a_{01}^-\nu t +\Ord\big(\nu^3,\nu t^2\big)\ep
\end{equation}

In order to derive the partial-wave HDRs, called Roy--Steiner (RS) equations, one needs to expand the $s$- and $t$-channel imaginary parts in~\eqref{eqn:HDRs} into the respective partial waves and subsequently project the full expanded equations onto either $s$- or $t$-channel partial waves; the resulting sets of integral equations together with the respective partial-wave unitarity relations then form the $s$- and $t$-channel RS subsystems.
According to~\cite{HiteSteiner}, the (unsubtracted) $s$-channel RS equations read (based on the MacDowell symmetry $f^I_{(\ell+1)-}(W)=-f^I_{\ell+}(-W)$ for all $\ell\geq0$~\cite{MacDowell})
\begin{align}
\label{eqn:sRS}
f^I_{\ell+}(W) &= N^I_{\ell+}(W)
+\frac{1}{\pi}\int\limits^\infty_{\tpi}\diff t'\sum\limits_J
\Big\{G_{\ell J}(W,t')\,\Im f^J_+(t')+H_{\ell J}(W,t')\,\Im f^J_-(t')\Big\}\nt\\
&\quad+\frac{1}{\pi}\int\limits^\infty_{W_+}\diff W'\sum\limits_{\ell'=0}^\infty
\Big\{K^I_{\ell\ell'}(W,W')\,\Im f^I_{\ell'+}(W')+K^I_{\ell\ell'}(W,-W')\,\Im f^I_{(\ell'+1)-}(W')\Big\}\ec
\end{align}
where due to $G$-parity only even/odd $J$ contribute for isospin $I=+/-$, respectively, and the partial-wave projections of the pole terms as well as the (lowest) kernels are analytically known, the latter including in particular the Cauchy kernel: $K^I_{\ell\ell'}(W,W')=\delta_{\ell\ell'}/(W'\!-W)+\dots$\,.
The $s$-channel $I=\pm$ partial waves are intertwined by the usual unitarity relations, which are diagonal in the $s$-channel isospin basis $I_s\in\{1/2,3/2\}$ only.
Once the $t$-channel partial waves are known, the structure of the $s$-channel RS subsystem is therefore similar to the \pipi Roy equations, cf.~\cite{ACGL}.
As shown in~\cite{DHKM}, the corresponding (unsubtracted) $t$-channel RS equations are given by
\begin{align}
\label{eqn:tRS}
f^J_+(t) &= \tilde N^J_+(t)
+\frac{1}{\pi}\int\limits^{\infty}_{W_+}\diff W'\sum\limits^\infty_{\ell=0}
\Big\{\tilde G_{J\ell}(t,W')\,\Im f^I_{\ell+}(W')+\tilde G_{J\ell}(t,-W')\,\Im f^I_{(\ell+1)-}(W')\Big\}\nt\\
&\quad+\frac{1}{\pi}\int\limits^{\infty}_{\tpi}\diff t'\sum\limits_{J'}
\Big\{\tilde K^1_{JJ'}(t,t')\,\Im f^{J'}_+(t')+\tilde K^2_{JJ'}(t,t')\,\Im f^{J'}_-(t')\Big\}
\end{align}
and similarly for the $f^J_-$ except for the fact that these do not receive contributions from the $f^J_+$.
Here, only even or odd $J'$ couple to even or odd $J$ (corresponding to $t$-channel isospin $I_t=0$ or $I_t=1$), respectively, and $\tilde K^1_{JJ'}$ (as well as the analogous $\tilde K^3_{JJ'}$ for the $f^J_-$) contains the Cauchy kernel.
Moreover, it turns out that only higher $t$-channel partial waves contribute to lower ones.
Assuming Mandelstam analyticity, the equations~\eqref{eqn:tRS} are valid for $\sqrt{t}\in[2\mpi,2.00\GeV]$ using $a=-2.71\mpis$, whereas~\eqref{eqn:sRS} holds for $W\in[\mN+\mpi,1.38\GeV]$ using $a=-23.19\mpis$.
The $t$-channel unitarity relations are diagonal in $I_t$ and only linear in the $f^J_\pm$ (below the first inelastic threshold $t_\tn{inel}$)
\begin{equation}
\Im f^J_{\pm}(t) = \sigma^\pi_t \big(t^{I_t}_J(t)\big)^* f^J_{\pm}(t)\,\theta\big(t-\tpi\big)\ec \qquad 
\sigma^\pi_t \, t^{I_t}_J(t)=\sin\delta^{I_t}_J(t)\,e^{i\delta^{I_t}_J(t)}\ec \qquad 
\sigma^\pi_t(t)=\sqrt{1-\tpi/t}\ec\nt
\end{equation}
from which one can infer Watson`s final state interaction theorem~\cite{Watson} stating that (in the ``elastic'' region) the phase of $f^J_\pm$ is given by the phase $\delta^{I_t}_J$ of the respective \pipi scattering partial wave $t^{I_t}_J$.

Due to the simpler recoupling scheme for the $f^J_\pm$, the $t$-channel RS subsystem can be recast as a (single-channel) Muskhelishvili--Omn\`es (MO) problem~\cite{Muskhelishvili+Omnes} with a finite matching point $\tm$~\cite{BDM} for $f^0_+$, $f^J_-$, and the linear combinations $\Gamma^J(t)=\mN\sqrt{J/(J+1)}\,f^J_-(t)-f^J_+(t)$ with $\Gamma^J(\tN)=0$ for all $J\geq1$ of the generic form (the details are given in~\cite{DHKM})
\begin{equation}
f(t) = 
\Delta(t)+\frac{1}{\pi}\int\limits_{\tpi}^{\tm}\diff t'\frac{\sin\delta(t')\,e^{-i\delta(t')}\,f(t')}{t'-t}
+\frac{1}{\pi}\int\limits_{\tm}^\infty\diff t'\frac{\Im f(t')}{t'-t}
\equiv\big|f(t)\big|e^{i\delta(t)}
\quad \tn{for }t\leq\tm\!<t_\tn{inel}\ec\nt
\end{equation}
where the inhomogeneities $\Delta(t)$ subsume the nucleon pole terms, all $s$-channel integrals, and the higher $t$-channel partial waves.
For $\tpi\leq t\leq\tm$, solving for $|f(t)|$ only according to Watson's theorem requires $\delta(t)$ for $\tpi\leq t\leq\tm$ and $\Im f(t)$ for $t\geq\tm$.
Introducing $n\geq1$ subtractions does not change the general structure of the RS/MO system, e.g.\ the $P$-waves are given by
\begin{equation}
\Gamma^1(t) = \Delta_\Gamma^1(t)\Big|^{n\tn{-sub}}\!\!
+\frac{t^{n-1}(t-\tN)}{\pi}\!\int\limits^\infty_{\tpi}\!\frac{\diff t'\,\Im\Gamma^1(t')}{t'^{n-1}(t'-\tN)(t'-t)}\ec \quad 
f^1_-(t) = \Delta^1_-(t)\Big|^{n\tn{-sub}}\!\!
+\frac{t^n}{\pi}\!\int\limits^\infty_{\tpi}\!\frac{\,\,\,\diff t'\,\Im f^1_-(t')}{t'^n(t'-t)}\ec\nt
\end{equation}
demonstrating that $\Gamma^J$ and hence $f^J_+$ is effectively subtracted by one power less than $f^J_-$, which motivates the additional (partial) third subtraction in $A^\pm$, cf.~\eqref{eqn:p3sub}, that affects solely the $f^J_+$.

\begin{figure}
\centering
\includegraphics[width=\textwidth]{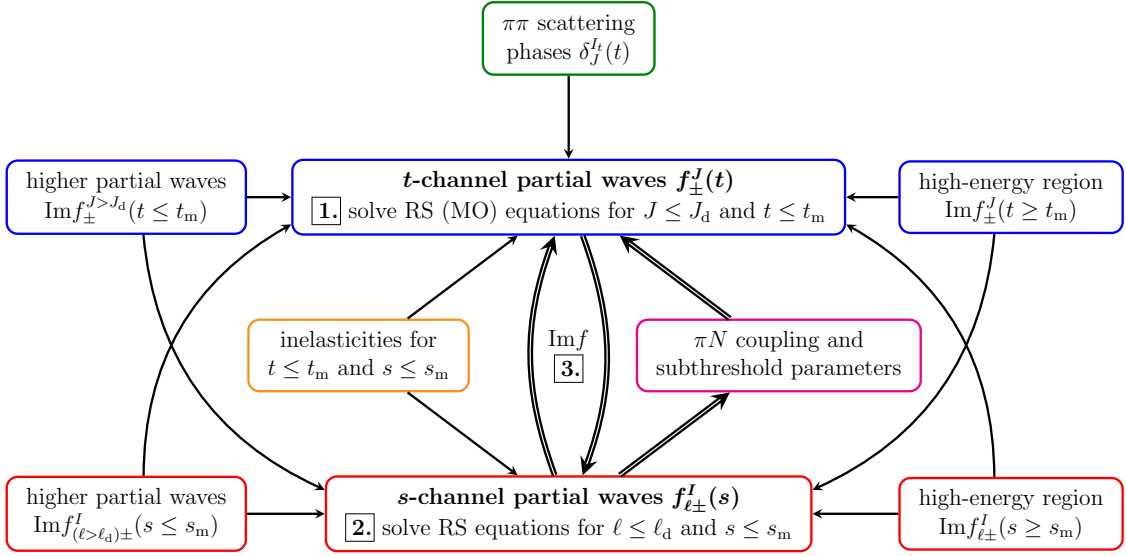}
\caption{Flowchart of the solution strategy for the Roy--Steiner system for \piN scattering. The third step consist in the self-consistent iteration (denoted by thick arrows) of the preceding steps until convergence.}
\label{fig:RSflowchart}
\end{figure}

The solution strategy for the full RS system in the low-energy (or even subthreshold/pseudo\-physical) regions, where only the lowest partial waves are relevant and inelastic contributions may be (approximately) neglected, is shown in Fig.~\ref{fig:RSflowchart}; see~\cite{DHKM} for more details.

\section{The $\boldsymbol{t}$-channel Muskhelishvili--Omn\`es problem: $\boldsymbol{P}$-wave solutions}

\begin{figure}
\centering
\includegraphics[width=0.49\textwidth]{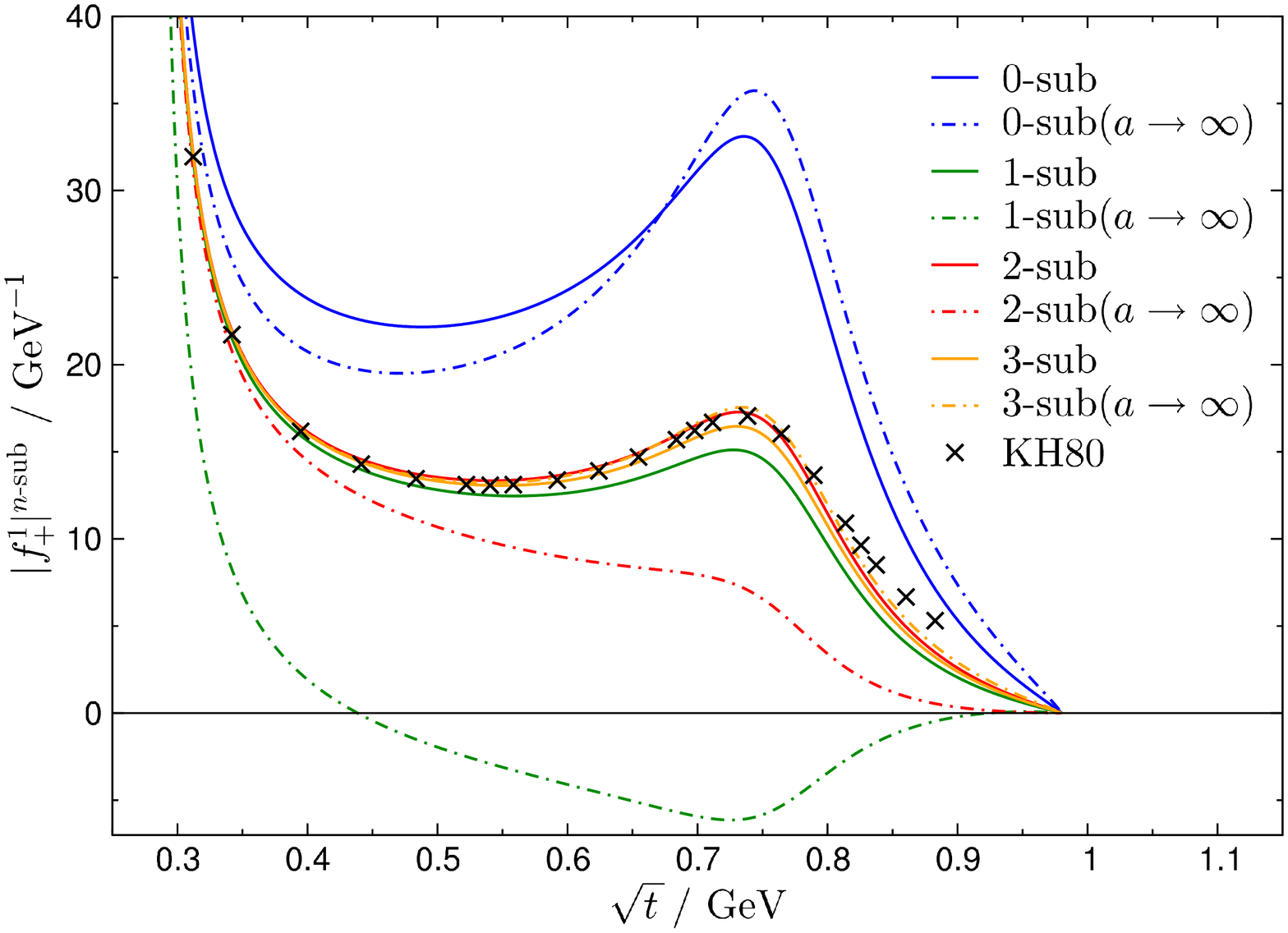}\hfill
\includegraphics[width=0.49\textwidth]{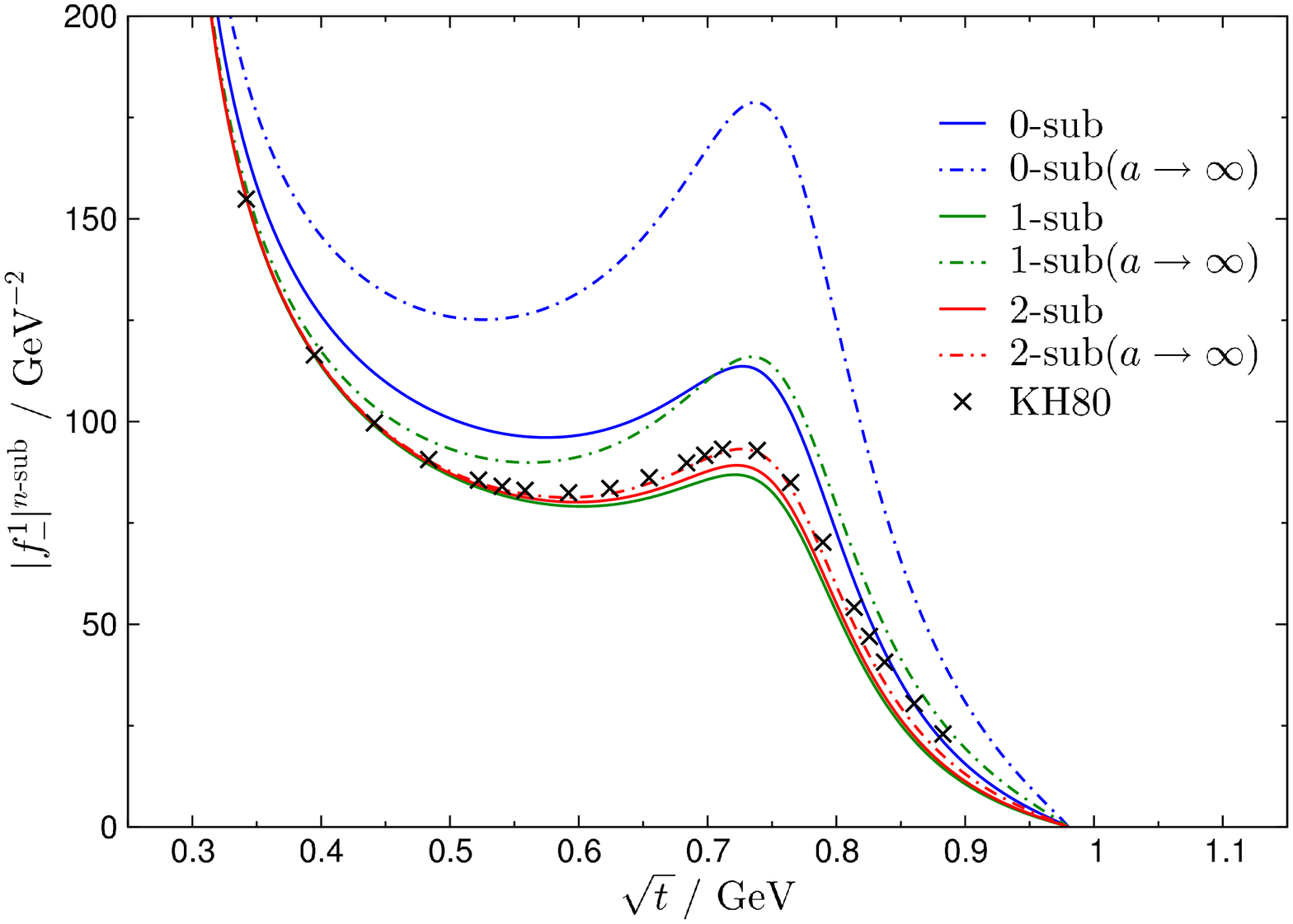}
\caption{$n$-subtracted MO solutions for the $P$-wave moduli.}
\label{fig:Pwaves}
\end{figure}

As the first step in the numerical solution of the full RS system, we check the consistency of our $t$-channel MO solutions with the results of the KH80 analysis~\cite{KH80}, which are still used nowadays although no thorough error estimates are given (and despite the availability of more modern experimental data).
Here, we present results for the $P$-waves in the (elastic) single-channel approximation of the MO problem, which is well justified for the $P$- and higher partial waves, whereas the $S$-wave requires a two-channel description including $\bar KK$ intermediate states as described in~\cite{DHKM}.
To produce the results (that will also serve as input for the solution of the $s$-channel RS subsystem, cf.\ Fig.~\ref{fig:RSflowchart}) partly shown in Fig.~\ref{fig:Pwaves}, we have used as input \pipi phase shifts from~\cite{CCL:Regge+PWA}, $s$-channel partial waves ($l\leq4$) from SAID~\cite{SAID+GWU:2008} for $W\leq2.5\GeV$, and above the Regge model of~\cite{Huang}.
To facilitate comparison with the results of KH80, we use the respective subthreshold parameter values and a \piN coupling of $g^2/(4\pi)=14.28$~\cite{Hoehler,KH80} (as starting point, the final values will result from the iteration procedure, cf.\ Fig.~\ref{fig:RSflowchart}).\footnote{Modern analyses yield significant smaller values for the \piN coupling, cf.\ e.g.\ $g^2/(4\pi)=13.7\pm0.2$ of~\cite{piNcoupling}.}
Moreover, KH80 uses different types of dispersion relations, in particular so-called fixed-$t$ ones, which can be emulated (up to the $t$-channel contributions that are not present at all in the fixed-$t$ case) by taking the ``fixed-$t$ limit'' $|a|\to\infty$.
As argued in~\cite{DHKM}, all $t$-channel input above $\sqrt{\tm}=0.98\GeV$ is set to zero, which forces the MO solutions to match zero at $t=\tm$.
While Fig.~\ref{fig:Pwaves} displays the results for $|a|\to\infty$, investigating the effect of using a different (i.e.\ higher) matching point leads to the same conclusion:
with increasing number of subtractions, thus lowering the dependence on the high-energy input by introducing more subthreshold parameter contributions as subtraction polynomials, the solutions show a nice convergence pattern both in general (proving the internal consistency and numerical stability of our RS/MO framework) and in particular towards the KH80 results (being consistent with relying on KH80 values for $g$ and the subtraction parameters).
\begin{figure}
\centering
\includegraphics[width=0.49\textwidth]{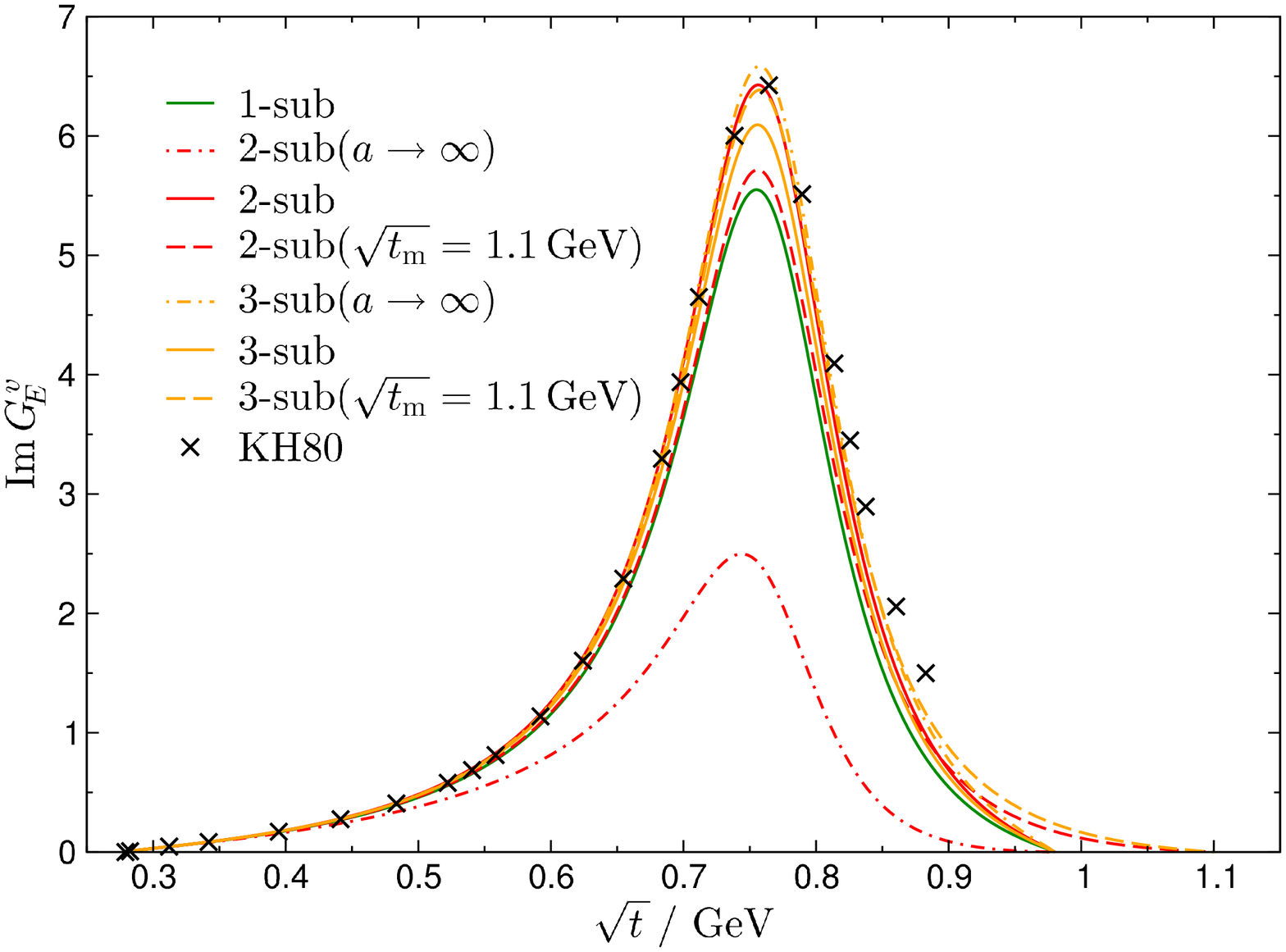}\hfill
\includegraphics[width=0.49\textwidth]{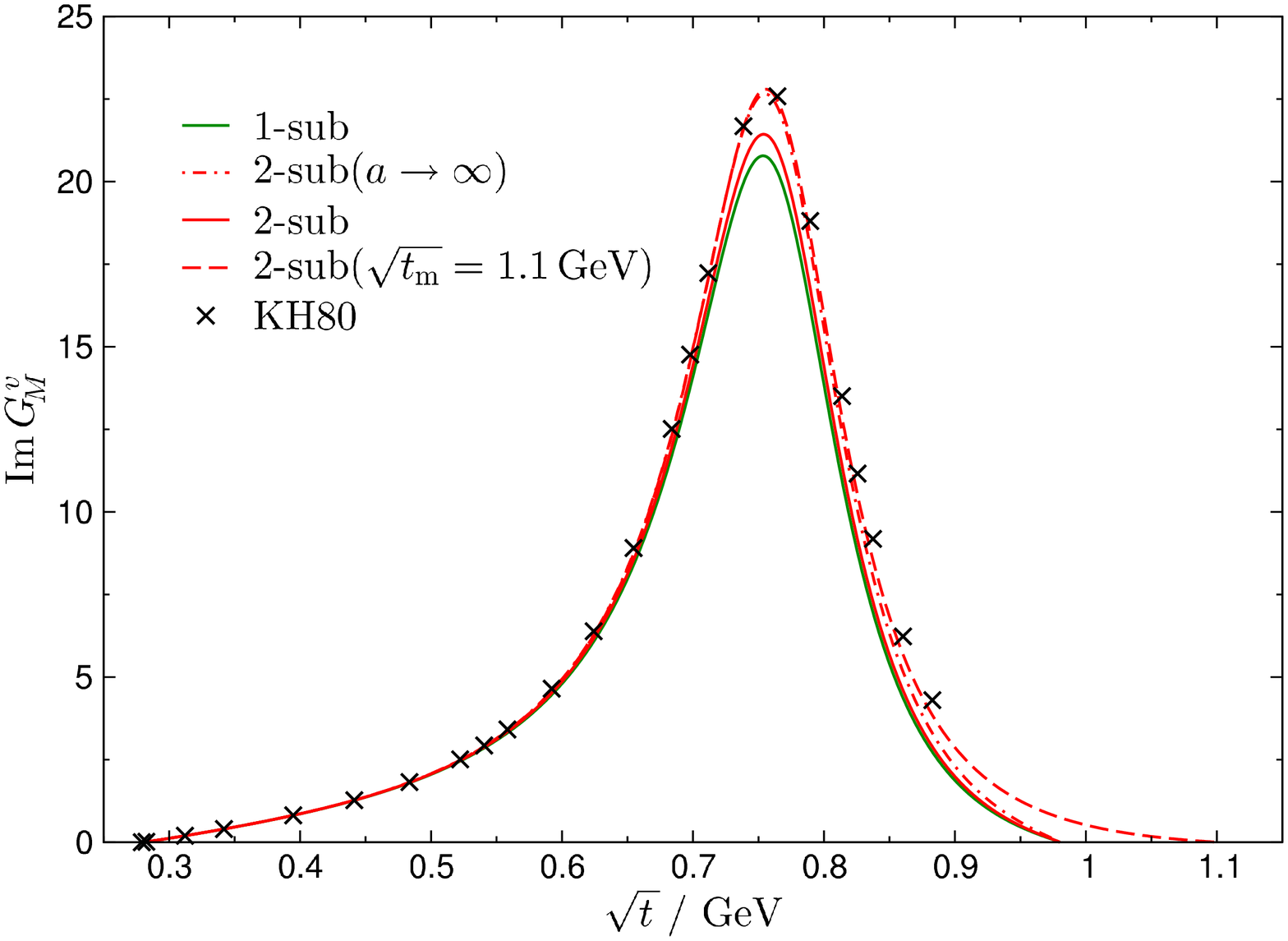}
\caption{Two-pion-continuum contribution to $\Im G_E^v$ and $\Im G_M^v$.}
\label{fig:ImGEM}
\end{figure}
The $P$-waves feature prominently in the dispersive analysis of the nucleon electromagnetic form factors, see e.g.~\cite{Lorenz} and references therein, and in Fig.~\ref{fig:ImGEM} we illustrate the effects on the spectral functions (by approximating the vector pion form factor $F_\pi^V$ via a twice-subtracted Omn\`es representation, cf.~\cite{DHKM})
\begin{equation}
\Im G_E^v(t)=\frac{t(\sigma^\pi_t)^3}{8\mN}\big(F_\pi^V(t)\big)^*f^1_+(t)\,\theta\big(t-\tpi\big)\ec \qquad 
\Im G_M^v(t)=\frac{t(\sigma^\pi_t)^3}{8\sqrt{2}}\big(F_\pi^V(t)\big)^*f^1_-(t)\,\theta\big(t-\tpi\big)\ep\nt
\end{equation}

We are confident that a self-consistent iteration procedure between the solutions for the $s$- and $t$-channel eventually will yield a consistent and precise description (including error estimates) of the low-energy \piN scattering amplitude in all kinematical channels.

\end{document}